**Title: Antarctic ice sheet – climate feedbacks under high future carbon emissions**

**Authors:** Shaina Rogstad[1*], Alan Condron[2], Robert DeConto[1], and David Pollard[3]

**Affiliations:**

[1] Climate System Research Center, Department of Geosciences, University of Massachusetts, Amherst, MA 01003.

[2] Department of Geology and Geophysics, Woods Hole Oceanographic Institution, Woods Hole, MA 02543, USA.

[3] Earth and Environmental Systems Institute, College of Earth and Mineral Sciences, Pennsylvania State University, PA 16802, USA.

*correspondence to: srogstad@geo.umass.edu

**One Sentence Summary:** Over the next 250 years, freshwater runoff from Antarctica could induce competing climate feedbacks via ocean heat, air temperatures, and sea ice.

**Abstract:** Freshwater forcing from a retreating Antarctic Ice Sheet could have a wide range of impacts on future global climate. Here, we report on multi-century (present-2250) climate simulations performed using a fully coupled numerical model integrated under future greenhouse gas emissions scenarios IPCC RCP4.5 and 8.5, with meltwater discharge provided by a dynamic-thermodynamic ice sheet model. Accounting for Antarctica's meltwater contribution raises sub-surface ocean temperatures by >1ºC at the ice margin, beyond rates current projected. In contrast, 2-10ºC cooler surface air and ocean temperatures in the Southern Ocean cause sea ice to expand and delay the increase of projected global mean anthropogenic warming. In addition, the projected loss of Arctic winter sea ice and weakening of the Atlantic Meridional Overturning Circulation are delayed be several decades. Our results demonstrate a clear need to more accurately account for meltwater input from ice sheets if we are to make confident climate predictions.

**Main Text:** Observational evidence indicates that the West Antarctic Ice Sheet (WAIS) is losing mass at an accelerating rate[1,2]. Recent advances in ice sheet modeling have improved our understanding of Antarctic Ice Sheet (AIS) evolution in response to anthropogenic greenhouse gas forcing and show that the AIS could contribute substantially to sea level rise by the end of this century[3–6]. A more accurate understanding of the impacts that this evolution might have on atmospheric and oceanic circulation dynamics is needed to constrain possible future changes in the climate system. However, ice sheet physics are not adequately represented in the current generation of global climate models (GCM) used in future projections[7,8]. The Antarctic Ice Sheet is considered a tipping element within the climate system[9], with the potential to contribute several tens of centimeters of global mean sea level rise in the next two centuries, but the climate system response to such large scale ice loss is not well constrained, especially beyond 2100.

Today, freshwater input to the ocean is increasing in response to climatic warming, largely from a combination of net precipitation and increasing riverine input resulting from an invigorated hydrologic cycle, and the loss of sea and land ice[10]. Previous modeling work investigating the relative impacts of freshwater forcing in the North Atlantic versus the Southern Ocean[11,12] has demonstrated that the location and magnitude of the additional freshwater are central to the modeled response. Methodology for modeling the climatic impact of freshwater perturbations has also varied widely in terms of strength, duration, and location of meltwater input: historically so-called 'hosing' approaches added water uniformly within given latitude bands[11–14], while more recent work has applied freshwater forcing at specific locations around global coastlines or spread according to iceberg movements[6,15–19]. Despite differences in model resolution and representation of earth system processes, several elements of the climate response to freshwater perturbations in the Southern Ocean have been broadly consistent, such as a decrease in surface air temperatures over the Southern Ocean, a decrease in the strength of the Atlantic Meridional Overturning Circulation (AMOC), and the expansion of sea-ice.

Here, we present results from a series of climate model simulations performed using a high-resolution, fully coupled, ocean-atmosphere-cryosphere-land model Community Earth System Model (CESM) 1.2.2 with CAM5 atmospheric physics[20] under Representative Concentration Pathways (RCP) 4.5 and 8.5[10] spanning 2005-2250 (See Methods). In our freshwater simulations, referred to throughout the paper as RCP4.5FW & RCP8.5FW, time-evolving freshwater runoff from Antarctica is provided from a continental ice sheet-ice shelf model[3] responding to the same atmospheric forcing scenarios, while the control runs (RCP4.5CTRL and RCP8.5CTRL) have no additional freshwater forcing beyond what is already simulated by the CESM model. To account for spatial and temporal variations in runoff, and to improve on classic 'hosing' experiments, time variant freshwater was released into the ocean at the nearest coastal grid cell to where ice calving or ocean melt is discharged in the ice-sheet (Fig. 1A; See Methods) such that considerable volumes of meltwater enter the ocean from the Amundsen coast of West Antarctica, including Pine Island and Thwaites glaciers. In RCP4.5FW, runoff increases throughout the 21$^{st}$ century, and remains between 0.4-0.8 Sv (Sv = 10$^6$ m$^3$/s) from 2050-2250; in contrast, the meltwater input in RCP4.5CTRL never exceeds 0.1 Sv (Fig 1B). In RCP8.5FW, freshwater input is dominated by the retreat of the WAIS in the ice sheet model during the 21$^{st}$ century, peaking at >2 Sv around ~2125. Runoff then remains above 1 Sv through 2200 due to increasing contributions from the East Antarctic Ice Sheet (EAIS). This is in sharp contrast to RCP8.5CTRL in which runoff increases steadily throughout the run, but never exceeds 0.2 Sv (Fig. 1D). As such, our methodology allows a direct comparison of the climate response to changing atmospheric greenhouse gas concentrations with and without a major Antarctic meltwater contribution, and allowing for partitioning between liquid and solid meltwater components (see Methods).

**Results**

The impact of applying a spatially varying freshwater forcing is immediately apparent in the salinity field. By the end of the 21$^{st}$ century, the sea surface salinity (SSS) in the RCP8.5FW experiment is reduced by up to -5 psu (compared to RCP8.5CTRL) over most of the Southern Ocean, and begins spreading northward (Fig. 1 and fig. S1). By the time of peak WAIS retreat, around 2120, the negative SSS anomaly exceeds -15 psu around the Antarctic margin, especially in the Amundsen and Bellingshausen Seas and portions of the Ross and Weddell Seas (Fig. 1C). By the middle of the 22$^{nd}$ century the anomaly has spread pervasively throughout all the ocean basins, to depths of ~4000 m (fig. S1). In RCP4.5FW, the ice sheet collapse does not peak and decline in the same way as RCP8.5FW, but rather is maintained throughout most of the run, resulting in the input of freshwater being persistent and steady (Fig. 1B, D). The associated salinity anomaly patterns are spatially similar to the RCP8.5FW simulation, but lower in magnitude (-1 to -2 psu) and remain confined to the Southern Ocean (fig. S2).

The prescribed changes in freshwater forcing from the ice sheet model have a profound impact on sea ice. Accurately capturing this response is important because seasonal freeze and melt cycles in the Southern Ocean act as a deepwater pump; thus, changes in sea ice are linked to changes in Southern Ocean overturning. The balance between brine rejection from sea-ice formation, freshwater forcing, and associated changes in ocean convection also lead to alterations in air-sea heat exchange which can trap warm waters at depths and increase melt rates under neighboring ice shelves[21]. Substantial changes in sea-ice extent impact the radiative balance through sea-ice albedo feedbacks and have broad implications for temperatures regionally in the Southern Ocean, as well as globally. In addition, changes in sea ice can dramatically impact a wide variety of biological species. For example, shifts in sea ice formation have already begun to impact penguin colonies and will likely have wide reaching effects on micro-fauna communities, krill abundance, and larger ocean predators[22,23].

In our simulations, sea-ice expands in both RCP4.5FW and RCP8.5FW, despite the elevated radiative forcing (Fig. 2). The large freshwater discharge in both simulations reduces salinity, raises the freezing temperature, and stratifies the water column around the coast. This in turn reduces convection, suppresses Southern Ocean overturning, and leads to a significant build-up in perennial sea-ice extent and thickness. The percent change in Southern Ocean area covered by sea-ice relative to modern varies over the course of the perturbation simulations from +100% to -85% (Table S1). Spatially, the greatest sea-ice growth in the perturbation experiments is within the South Pacific sector, where the freshwater input is largest. In RCP8.5FW, Southern Ocean sea-ice extent reaches a maximum in the 2120's during peak freshwater discharge with sea-ice thickness exceeding 10 m in the Amundsen, Bellingshausen, and Ross Seas, and parts of the EAIS margin (Fig. 2). As freshwater forcing declines following WAIS collapse, sea-ice extent and thickness also begin to decline, although sea-ice >10 m thick still persists in several regions in year 2200 (fig. S3a, e). Sea ice accumulates within the first few decades in the RCP4.5FW run, but after the 2120's, when peak freshwater discharge has occurred in RCP8.5FW, sea ice extent and thickness are significantly more extensive in RCP4.5FW where perennial sea ice >5 m thick persists into the 22$^{nd}$ century despite the anthropogenic greenhouse gas forcing (fig. S3c, g). In contrast to the large quantities of sea ice produced in the perturbation experiments, sea-ice never expands in RCP4.5CTRL and RCP8.5CTRL and declines over the course of those runs with minimal sea-ice in the Southern Ocean by 2100, and no austral winter sea-ice by 2200 (Fig. 2A and fig. 4).

The effect of Antarctic freshwater forcing on sea-ice produces a strong albedo feedback contributing to delayed atmospheric warming in the perturbation experiments (Fig. 1B, D). Spatially, the cooler temperatures relative to the control simulations are maximized directly over the Antarctic continental margin where the freshwater perturbation is applied (Fig. 3A, B). The effect of the freshwater forcing on global mean surface temperature (GMST) reaches a maximum at the time of peak ice sheet retreat in RCP8.5FW, with GMST values 2.5°C lower than the control run (Fig. 1b and fig. S4). This finding demonstrates that Antarctic ice loss could provide a negative feedback on anthropogenic warming, despite catastrophic impacts to the climate system as a whole and substantial contributions to sea level rise. While the anthropogenic warming is mitigated somewhat until Antarctica is exhausted of ice under RCP8.5FW, global temperatures still rise substantially over the course of the simulations. (Fig. 1b and Table S1).

Antarctic freshwater forcing strongly modifies the trajectory of polar climate in both hemispheres. During peak WAIS collapse, when the surface air temperature in the Arctic (north of 60°N) is up to 2.5°C cooler in RCP8.5FW compared to RCP85CTRL, the decline in Arctic winter sea-ice is slowed such that complete loss of Arctic sea ice is delayed by ~30 years (fig. S5). In the Southern Ocean, expanded sea-ice growth suppresses warming, particularly in the Amundsen Sea region of Antarctica where sea-ice formation is maximized. The freshwater-sea ice cooling feedback is so strong that surface air temperatures in portions of the Southern Ocean are colder after 2100 than at the beginning of the simulation in the early 21st century. This effect is seen in both RCP4.5FW and RCP8.5FW. It persists until the end of run under RCP4.5FW as mass loss is slower, however disappears in RCP8.5FW after the peak of freshwater forcing when the West and East Antarctic basins become exhausted of ice and temperatures over the Southern Ocean begin to rise rapidly, ending >10°C warmer than the start of the run (fig. S6).

Global sea surface temperatures (SST) increase due to anthropogenic emissions throughout all simulations. During RCP8.5FW, the perturbation region in the Southern Ocean is an exception as SSTs cool by as much as 2°C during the first century and through the period of peak freshwater discharge, as compared to the start of the run, despite the radiative forcing changes (fig. S7). Compared to RCP8.5CTRL, we find that SSTs in RCP8.5FW are significantly different with a 2-10°C cooling in the Southern Hemisphere at the time of peak freshwater forcing during the 2120's, while a slight warming of ~2°C is observed in the North Atlantic and sub-tropical Pacific (Fig. 3C). The spatial patterns of temperature change between the RCP4.5FW and RCP4.5CTRL simulations are similar to those in RCP8.5, but of smaller magnitude. For example, SSTs in the Southern Hemisphere are 1-3°C cooler, while in the North Atlantic and sub-tropical Pacific the warming is, at most, ~0.5-1°C (Fig. 3D).

The cooling response of the Southern Ocean surface contrasts with a subsurface (~400 m, representative of continental shelf depths at the mouth of ice-shelf cavities) warming around the ice sheet margin. This juxtaposition is caused by the expanded sea-ice cover, intense surface stratification in the water column, and reduced vertical mixing as seen in other studies[18]. The effect is more intense in our simulations because our integrations are run forward long enough to capture the peak RCP8.5 freshwater discharge associated with maximum WAIS retreat in the ice sheet model in the early 22$^{nd}$ century. The strongest subsurface ocean warming in RCP8.5FW is seen in the Ross Sea, with temperatures at 400 m water depth 2-4° C warmer than RCP8.5CTRL in the 2120s (Fig. 3E). The strongest warming in our RCP4.5FW experiment is observed in the Weddell Sea at this time (Fig. 3F), although as noted previously the WAIS does not undergo the same rapid collapse in this scenario. By 2250, temperatures are up to 3°C warmer in RCP4.5FW and up to 6°C warmer in RCP8.5FW as compared to the start of run averages (fig. S8). The subsurface warming effect remains confined to the Southern Ocean south of the Antarctic Circumpolar Current, as large parts of the deep ocean display the same cooling anomaly seen in the SSTs (fig. S9).

Past changes in overturning strength are associated with rapid shifts in past climate[24]. Observational records show the Atlantic Meridional Overturning Circulation (AMOC) has slowed since the 1950's[25]. In previous Southern Ocean freshwater forcing experiments[11,14], a low-salinity anomaly was found to spread northward into the North Atlantic, suppressing deepwater formation. However, those experiments applied the freshwater forcing uniformly over a large region of the Southern Ocean rather than at the location of ice and meltwater discharge around the Antarctic margin. In our experiments, the salinity anomaly spreads throughout the Southern Ocean, but it does not reach the North Atlantic at sufficient strength to inhibit overturning. This difference could be a result of the salinity perturbation in these earlier studies being applied across the Southern Ocean, rather than only in grid cells adjacent to the ice sheet as in this study[26].

To assess the impact of Antarctic freshwater forcing on overturning strength, the maximum overturning values throughout the full depth range of the water column were calculated in the Atlantic Ocean from 20-50°N. In both RCP8.5 simulations an almost complete collapse of the overturning circulation is seen with the strength of the overturning circulation decreasing from 24 Sv in 2005 to 8 Sv by 2250 (Fig. 4A). However, the freshwater perturbation in RCP8.5FW delays the collapse of the overturning circulation by 35 years relative to RCP8.5CTRL, based on the timing when overturning strength drops below 10 Sv for 5 consecutive years. The maximum effect of ice-sheet freshwater forcing on overturning strength corresponds with peak ice sheet retreat around 2120. The relatively greater AMOC strength seen in RCP8.5FW may be a contributing factor to the higher SST and SAT temperatures seen in the North Atlantic at this time as compared to RCP8.5CTRL. In RCP4.5FW, the strength of the overturning declines in the beginning of the run and settles into a lower equilibrium of 19 Sv, but does not fully collapse. The decline is delayed between 2050-2100 compared to RCP4.5CTRL. After 2200, AMOC begins to recover in RCP4.5CTRL, but remains suppressed in RCP4.5FW (Fig. 4A).

Meltwater-forced changes in the overturning circulation impacts northward heat transport in the Atlantic Ocean (Fig. 4C). In our RCP8.5FW experiment, we find that during the period of maximum freshwater forcing the largest anomaly in northward heat transport (compared to RCP8.5CTRL) is between 20-40ºN with a difference of ~0.16 PW (1 PW=$10^{15}$ watts). A similar pattern emerges for the RCP4.5 anomaly, but to a lesser extent. Finally, the delayed warming in the Southern Hemisphere and enhanced warming in the North Hemisphere, along with a stronger AMOC in the meltwater perturbation runs, results in a northward shift in the Inter Tropical Convergence Zone (ITCZ) under both RCP4.5FW and RCP8.5FW scenarios.

The patterns of precipitation change are broadly similar in both experiments, although the magnitude of the changes are generally smaller in the RCP4.5FW scenario (Fig. 4 B,D).

**Conclusion**

In summary, our climate model simulations show that future changes in meltwater runoff from the AIS will have major implications for both regional and global climate. The multi-century simulations for RCP8.5 shown here, spanning the interval of peak freshwater discharge in the 22$^{nd}$ century, accounting for the spatial variability in freshwater distribution, and partitioning of liquid and solid water discharge simulated by an ice sheet model[3], highlight a range of feedback mechanisms that demonstrate the importance of freshwater forcing on global climate. Our results also point to a more complicated picture of WAIS stability based on standalone ice-sheet simulations, without accounting for ice-ocean-atmosphere interactions. By including Antarctic freshwater forcing in future greenhouse gas forcing scenarios we find that the increased stratification of the Southern Ocean and the large-scale build-up of sea-ice causes subsurface warming that can accelerate sub-ice melt rates and ice shelf loss. At the same time, increased albedo from Southern Ocean sea-ice growth provides a strong negative feedback on air surface temperatures over the continental margin that could mitigate surface melt and hydrofracturing of ice shelves. Finally, we find a delay in the future decline in North Atlantic overturning strength that enhances northward heat transport.  A previous study attempting to capture meltwater feedbacks on ice sheet simulations using an intermediate-complexity climate model[6], found the subsurface ocean warming feedback dominates over changes in surface air temperatures, but this ice sheet model did not account for hydrofracturing[3], so the relative importance of these competing feedbacks (subsurface ocean warming versus atmospheric cooling) have yet to be fully tested. The results shown here clearly demonstrate the need for interactive, or fully synchronous, simulations of ice sheets with fully coupled global climate models, to more accurately assess the future stability of the Antarctic Ice Sheet and the broader global climate impacts of significant ice loss from Antarctica[6].

**References and Notes**

1. Konrad, H. *et al.* Net retreat of Antarctic glacier grounding lines. *Nat. Geosci.* **11**, 258 (2018).
2. Mass balance of the Antarctic Ice Sheet from 1992 to 2017. *Nature* **558**, 219 (2018).
3. DeConto, R. M. & Pollard, D. Contribution of Antarctica to past and future sea-level rise. *Nature* **531**, 591–597 (2016).
4. Golledge, N. R. *et al.* The multi-millennial Antarctic commitment to future sea-level rise. *Nature* **526**, 421–425 (2015).
5. Ritz, C. *et al.* Potential sea-level rise from Antarctic ice-sheet instability constrained by observations. *Nature* **528**, 115–118 (2015).
6. Golledge, N. R. *et al.* Global environmental consequences of twenty-first-century ice-sheet melt. *Nature* **566**, 65 (2019).
7. Meijers A. J. S. The Southern Ocean in the Coupled Model Intercomparison Project phase 5. *Philos. Trans. R. Soc. Math. Phys. Eng. Sci.* **372**, 20130296 (2014).
8. Turner John, Bracegirdle Thomas J., Phillips Tony, J. Marshall Gareth & Scott Hosking J. An Initial Assessment of Antarctic Sea Ice Extent in the CMIP5 Models. *J. Clim.* 1473 (2013).
9. Lenton, T. M. *et al.* Tipping elements in the Earth's climate system. *Proc. Natl. Acad. Sci.* **105**, 1786–1793 (2008).
10. AR5 Climate Change 2013: The Physical Science Basis — IPCC.
11. Stouffer Ronald J., Seidov Dan & Haupt Bernd J. Climate Response to External Sources of Freshwater : North Atlantic versus the Southern Ocean. *J. Clim.* 436 (2007).
12. Ma Hao & Wu Lixin. Global Teleconnections in Response to Freshening over the

Antarctic Ocean. *J. Clim.* 1071 (2011).

13. Seidov, D., Barron, E. & Haupt, B. J. Meltwater and the global ocean conveyor: northern versus southern connections. *Glob. Planet. Change* **30**, 257–270 (2001).

14. Swingedouw, D., Fichefet, T., Goosse, H. & Loutre, M. F. Impact of transient freshwater releases in the Southern Ocean on the AMOC and climate. *Clim. Dyn.* 365 (2009).

15. Menviel, L., Timmermann, A., Timm, O. E. & Mouchet, A. Climate and biogeochemical response to a rapid melting of the West Antarctic Ice Sheet during interglacials and implications for future climate. *Paleoceanography* **25**, (2010).

16. Fogwill, C. J., Phipps, S. J., Turney, C. S. M. & Golledge, N. R. Sensitivity of the Southern Ocean to enhanced regional Antarctic ice sheet meltwater input. *Earths Future* **3**, 317–329 (2015).

17. Pauling, A. G., Bitz, C. M., Smith, I. J. & Langhorne, P. J. The Response of the Southern Ocean and Antarctic Sea Ice to Freshwater from Ice Shelves in an Earth System Model. *J. Clim.* **29**, 1655–1672 (2016).

18. Bronselaer, B. *et al.* Change in future climate due to Antarctic meltwater. *Nature* **564**, 53 (2018).

19. Schloesser, Fabian. Antarctic Iceberg impacts on future Southern Hemisphere Climate. *Reveiw*

20. Hurrell, J. W. *et al.* THE COMMUNITY EARTH SYSTEM MODEL. 22 (2013).

21. Merino, N. *et al.* Impact of increasing antarctic glacial freshwater release on regional sea-ice cover in the Southern Ocean. *Ocean Model.* **121**, 76–89 (2018).

22. Fretwell, P. T. & Trathan, P. N. Emperors on thin ice: three years of breeding failure at Halley Bay. *Antarct. Sci.* **31**, 133–138 (2019).

23. Chown, S. L. *et al.* The changing form of Antarctic biodiversity. *Nature* **522**, 431–438 (2015).

24. Rahmstorf, S. Ocean circulation and climate during the past 120,000 years. *Nature* **419**, 207–214 (2002).

25. Caesar, L., Rahmstorf, S., Robinson, A., Feulner, G. & Saba, V. Observed fingerprint of a

weakening Atlantic Ocean overturning circulation. *Nature* **556**, 191 (2018).

26. Condron, A. & Winsor, P. Meltwater routing and the Younger Dryas. *Proc. Natl. Acad. Sci.* **109**, 19928–19933 (2012).
27. Bintanja, R., Oldenborgh, G. J. van & Katsman, C. A. The effect of increased fresh water from Antarctic ice shelves on future trends in Antarctic sea ice. *Ann. Glaciol.* **56**, 120–126 (2015)., 120–126 (2015).
28. T. Fan, C. Deser, D. P. Schneider, Recent Antarctic sea ice trends in the context of Southern Ocean surface climate variations since 1950. Geophys. Res. Lett. 41, 2419–2426 (2014).
29. N. Merino, N. C. Jourdain, J. Le Sommer, H. Goosse, P. Mathiot, G. Durand, Impact of increasing antarctic glacial freshwater release on regional sea-ice cover in the Southern Ocean. Ocean Modelling. 121, 76–89 (2018).

**Acknowledgements:** We would like to thank Christine Shields and Nan Rosenbloom at NCAR for assistance with CESM. We would also like to thank Jan Lenaerts and Leo van Kampenhout for correspondence on freshwater forcing methodology. Preliminary numerical simulations to test methodology were performed in 2014-2016 using the high-performance computing support from Yellowstone provided by NCAR's Computational and Information Systems Laboratory, sponsored by the National Science Foundation. The final simulations presented here were performed with high-performance computing support from Cheyenne (*doi:10.5065/D6RX99HX*), also provided by the NSF funded NCAR Computational and Information Systems Laboratory. Data for the RCP4.5 control run was obtained through the Earth System Grid. **Funding:** This research was supported by the National Science Foundation Office of Polar Programs through NSF grant 1443347 and the Biological and Environmental Research (BER) division of the U.S. Department of Energy through grant DE-SC0019263. **Author contributions:** SR performed all the numerical simulations with help from AC, RD, and DP, and analyzed all the data. AC and RD conceived the original idea for the project and obtained funding through the National Science Foundation. SR wrote the manuscript with input from AC, RD, and DP. **Competing interests:** The authors declare no competing interests. **Data and methods availability:** The CESM model code is publicly available from NCAR. The results from the standard- and meltwater-ensemble simulations were archived at both Woods Hole Oceanographic Institution and UMass Amherst and are available from the corresponding author.

**Supplementary Materials:**

Materials and Methods

Figures S1-S9

Table S1

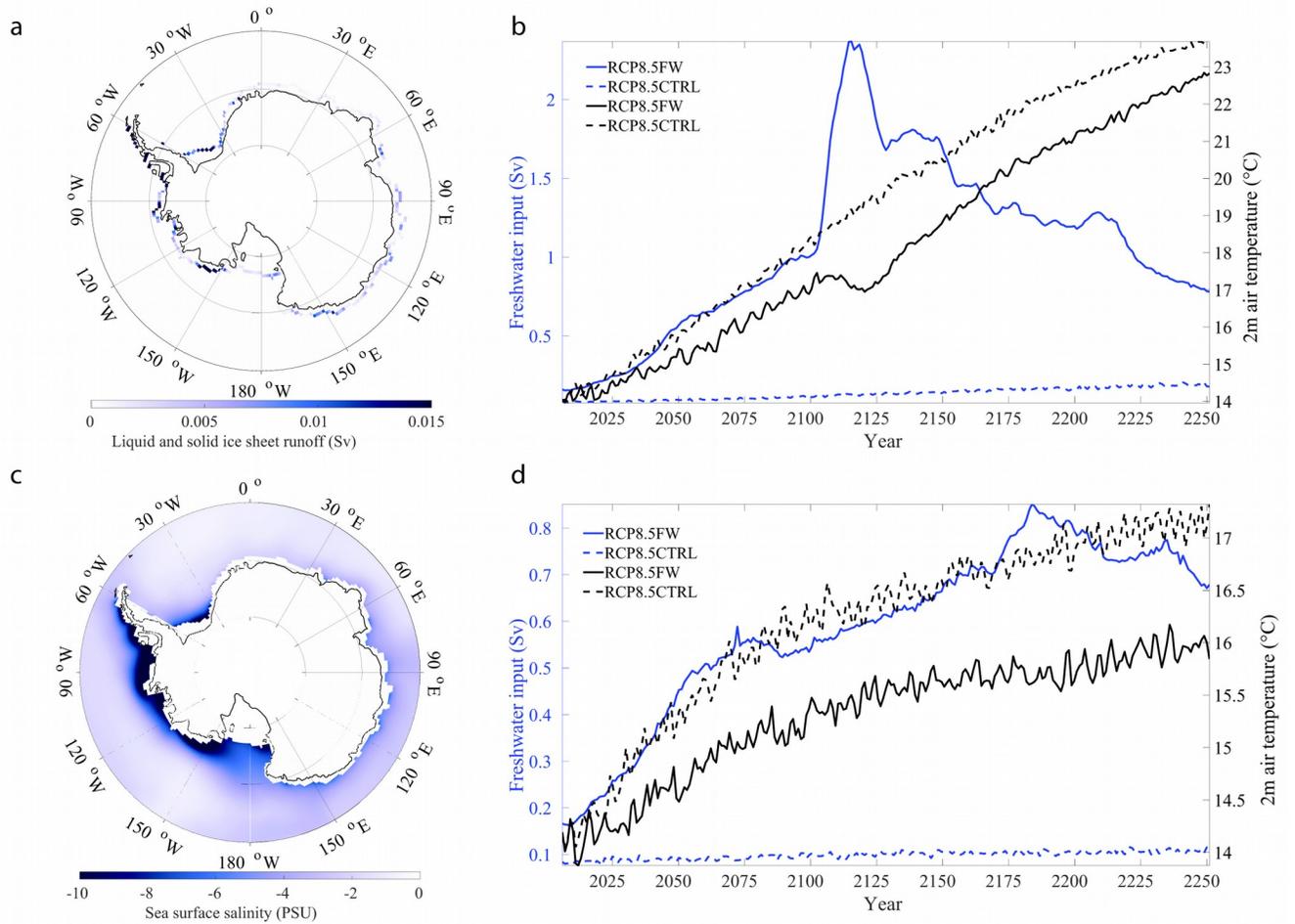

**Fig. 1. Freshwater forcing quantities and salinity response.** (**A**) Spatially distributed, time-varying freshwater forcing around the continental margin in September 2121 shown as the combination of liquid and solid input. (**B**) Freshwater forcing in relation to the global mean surface temperature in RCP8.5. (**C**) Decadal (2121-2130) sea-surface salinity difference between RCP8.5FW and RCP8.5CTRL, reflecting the freshwater input during peak ice sheet retreat. (**D**) The same as (B) except for RCP4.5.

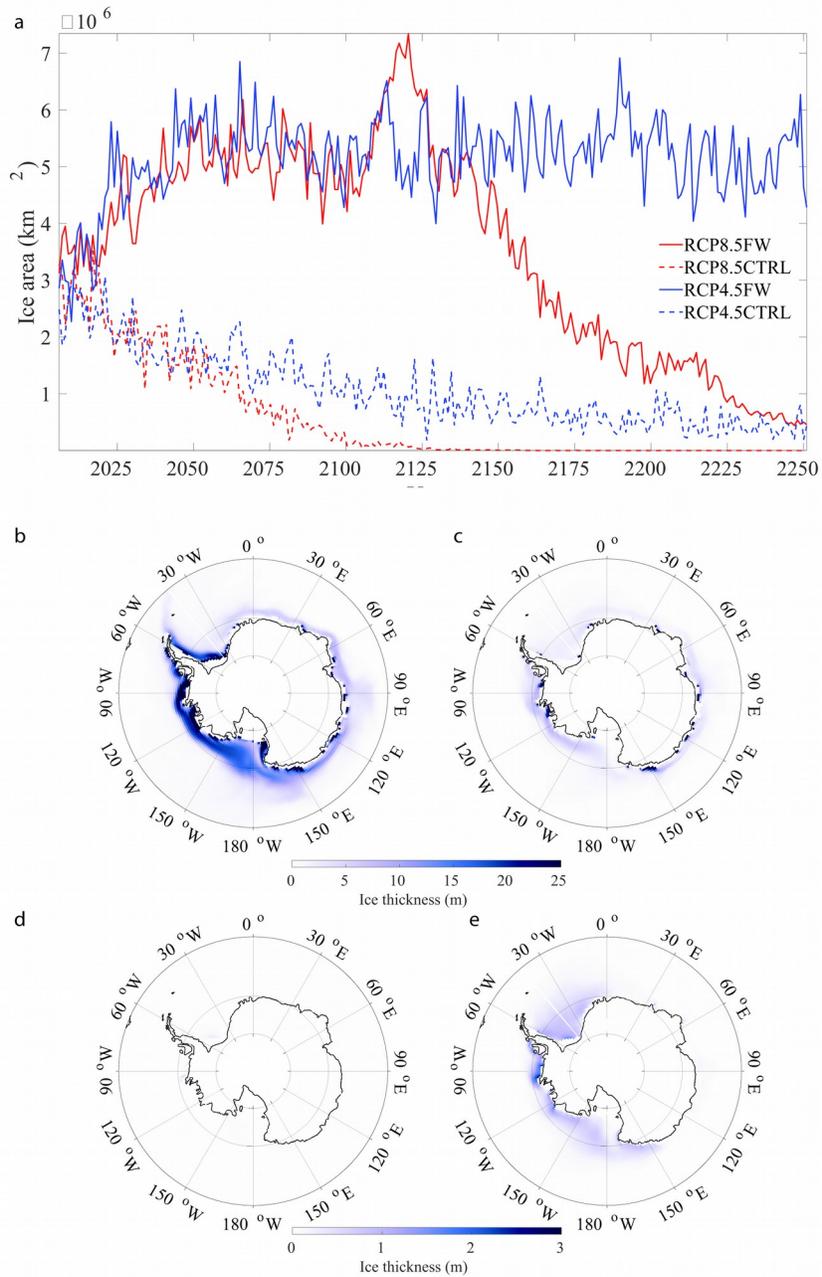

**Fig. 2. Sea ice response to freshwater forcing.** (**A**) Time series of Southern Ocean ice area in February showing the extent of perennial sea ice in austral summer. Lower anthropogenic radiative forcing allows for a much greater sea-ice area in the 22nd century in RCP4.5FW despite a similar magnitude of freshwater forcing to that of RCP8.5FW. (**B** to **E**) February sea ice thickness decadally averaged for 2121-2130 for RCP8.5FW (B), RCP4.5FW (C), RCP8.5CTRL (D), RCP4.5CTRL (E). Note the difference in scale for (D) to (E).

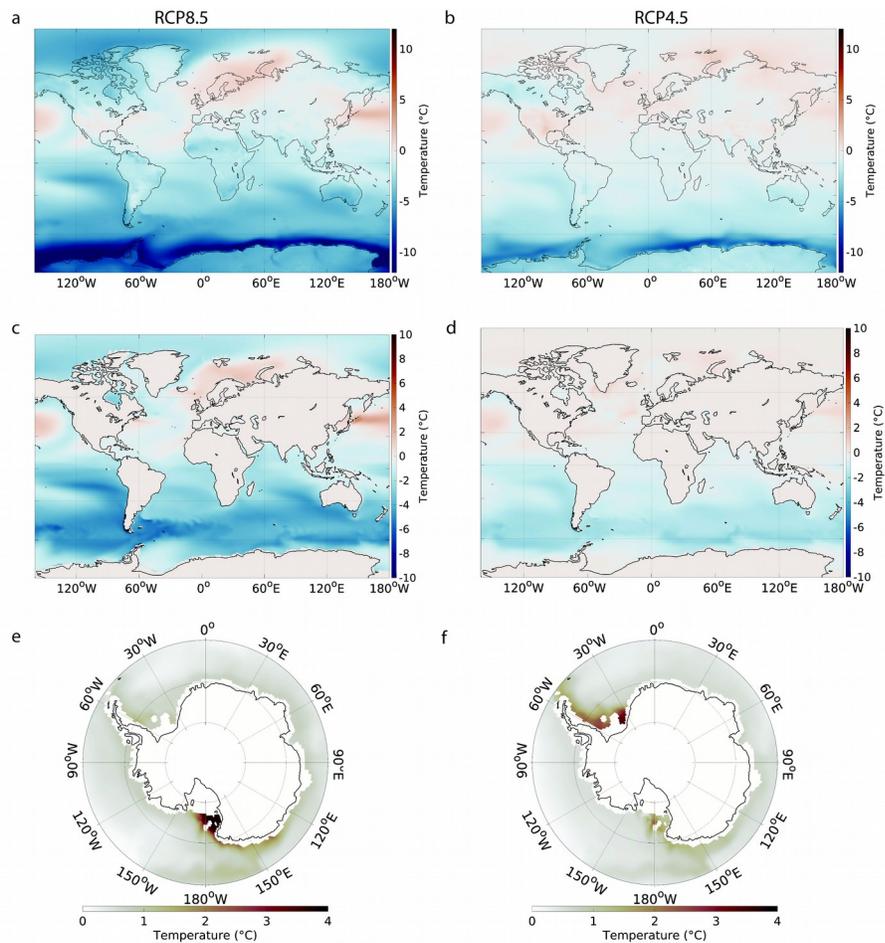

**Fig. 3. Air and ocean temperatures.** (**A**) Surface air temperature (SAT) difference (RCP8.5FW minus RCP8.5CTRL), decadally averaged for 2121-2130, show strong cooling throughout the Southern Ocean. (**B**) Same as (A), but for RCP4.5FW minus RCP4.5CTRL. Note the cooling is limited to the southern hemisphere. (**C**) Decadally averaged sea surface temperature (SST) difference (RCP8.5FW minus RCP8.5CTRL) for 2121-2130 showing Southern Ocean cooling spreading to the equator and parts of the northern hemisphere. (**D**) same as (C) except for RCP4.5FW minus RCP4.5CTRL. (**E**) Subsurface ocean temperature difference (RCP8.5FW minus RCP8.5CTRL) at 400 m water depth, representative of continental shelf depths at the mouth of ice-shelf cavities. Warming is concentrated in the Ross Sea. (**F**) Same as (E), but for RCP4.5FW minus RCP4.5CTRL, showing warming concentrated in the Weddell Sea.

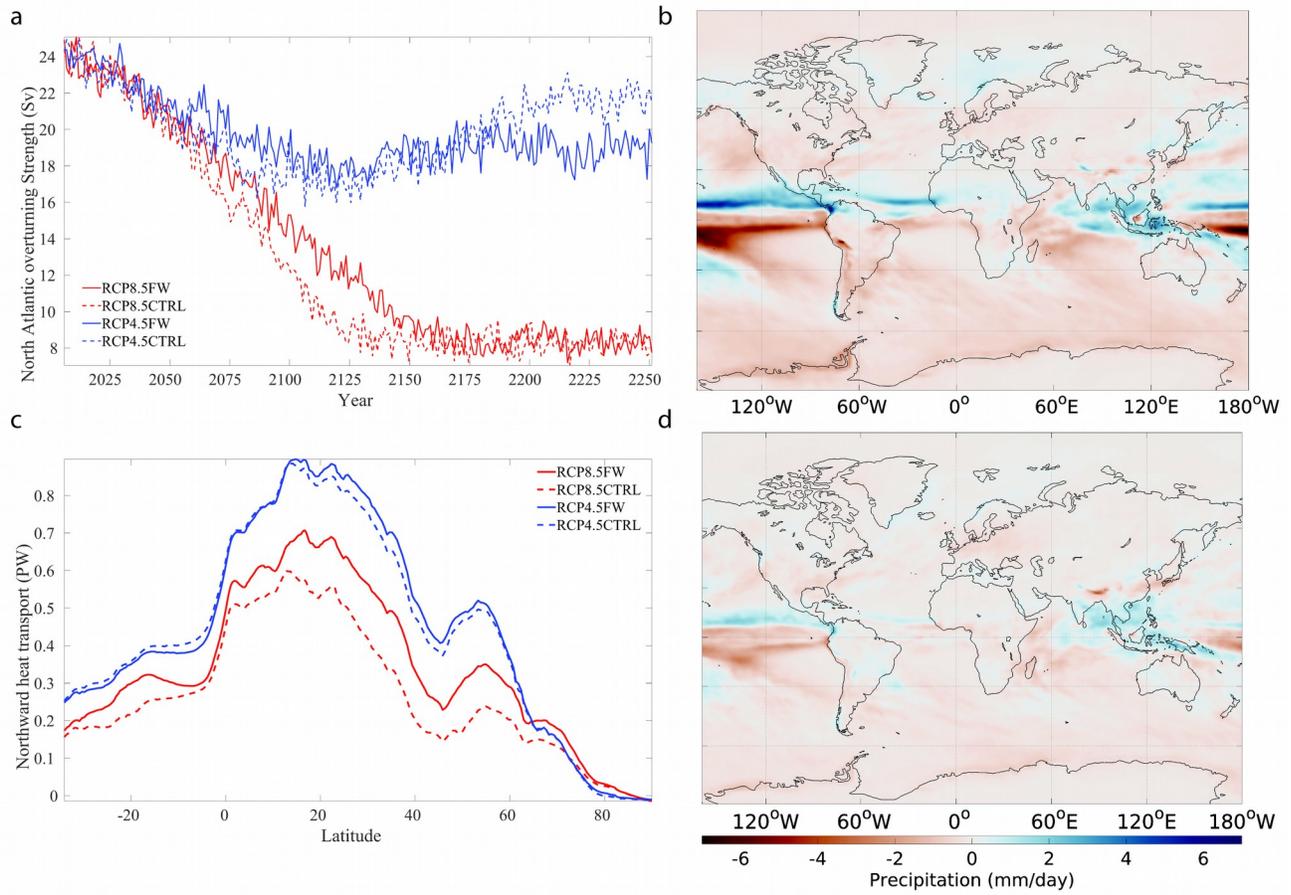

**Fig. 4. North Atlantic ocean heat transport, meridional overturning circulation and global precipitation.** (**A**) Time series of the Atlantic Meridional Overturning (AMOC) strength in Sv. (**B**) Decadally averaged precipitation difference for 2121-2130 (RCP8.5FW minus RCP8.5CTRL). (**C**) The northward heat transport difference for 2121-2130 (RCP8.5FW minus RCP8.5CTRL). (**D**) Same as (B) except for RCP4.5FW minus RCP4.5CTRL.

**Supplemental Material**

**Materials and Methods:**

Three model simulations were conducted using Community Earth System Model (CESM) 1.2.2 with CAM5 physics[23]. Model integrations were conducted using a 1° grid resolution for the ocean and sea-ice components, with a displaced pole over Greenland, and a finite volume 0.9x1.25° grid for the atmosphere and land components. The ocean model contains 60 vertical layers, and there are 30 vertical layers representing the atmosphere. Integrations were initialized from 20th century restart files and run under IPCC RCP4.5 and 8.5 greenhouse gas forcing scenarios from 2005-2250.

For the RCP4.5 and 8.5 perturbation simulations (RCP4.5FW & RCP8.5FW), the freshwater forcing data were obtained from previous offline ice-sheet model simulations, driven by the same RCP4.5 and 8.5 emissions scenarios[3]. In our CESM simulations, freshwater input from the Antarctic Ice Sheet is spatially and temporally distributed and differentiates between liquid and solid components (fig. S2). Partitioning of liquid and solid components within CESM has the advantage of taking into account latent heat of melting for the solid component. Accounting for latent heat has been found to be an important component in ocean response[19]. Liquid components from the ice sheet model include sub-ice ocean melt, cliff face melt, and percolation of meltwater to the base, while solid components represent ice calving and basal refreezing. Using the freshwater component quantities allows for a larger magnitude of input as opposed to using ice sheet volume change as done in previous studies[18]. The freshwater flux from the polar stereographic ice sheet model grid is spatially interpolated and applied as a surface freshwater perturbation at the nearest coastal grid cells following each longitude band in the CESM gx1v6 grid. This inputs freshwater at 320 grid cell locations around the continental margin. For the RCP8.5 control run (RCP8.5CTRL), freshwater runoff is calculated by the standard CESM with no additional freshwater forcing from the ice sheet model. Due to computational limitations, no control run was done for RCP4.5, and instead the data from the CCSM4 b.e11.BRCP45C5CN.f09_g16.001 run was obtained from Earth System Grid and used as a control (referred to as RCP4.5CTRL).

Recent observations show a northward expansion of sea-ice in some sectors of the Southern Ocean as well as a cooling of the ocean surface[28]. However, models from phase 5 of the Coupled Model Intercomparison Project (CMIP5) predict a sea-ice decline over the modern period continuing into the future[8]. Since freshwater forcing from the ice sheets is lacking in the current suite of climate models, inaccurate freshwater runoff has been suggested as the cause of discrepancies between models and observations[8]. Previous climate simulations using CESM1(CAM5) for 1980-2013[17] found that after an initial adjustment period, sea-ice area showed no increase in response to freshwater forcing, suggesting other methods could be at play in driving recently observed sea-ice trends. The freshwater forcing applied in Ref.[17] was much less than applied in our long-term future simulations. Modeling studies of future climate response to freshwater forcing in the Southern Ocean show expansion of sea-ice extent in response to freshwater perturbations[18,27]. There may be a threshold beyond which freshwater becomes a dominant control on sea-ice formation. Their study found that sea-ice response was insensitive to the perturbation depth where the freshwater was added to the ocean. Our study utilized a forcing scheme similar to that recently used by Ref[18], with freshwater applied at the surface only. Other groups have shown distinct regional differences in sea-ice sensitivity suggesting that regional differences in freshwater perturbations will be important for assessing future ice response[29].

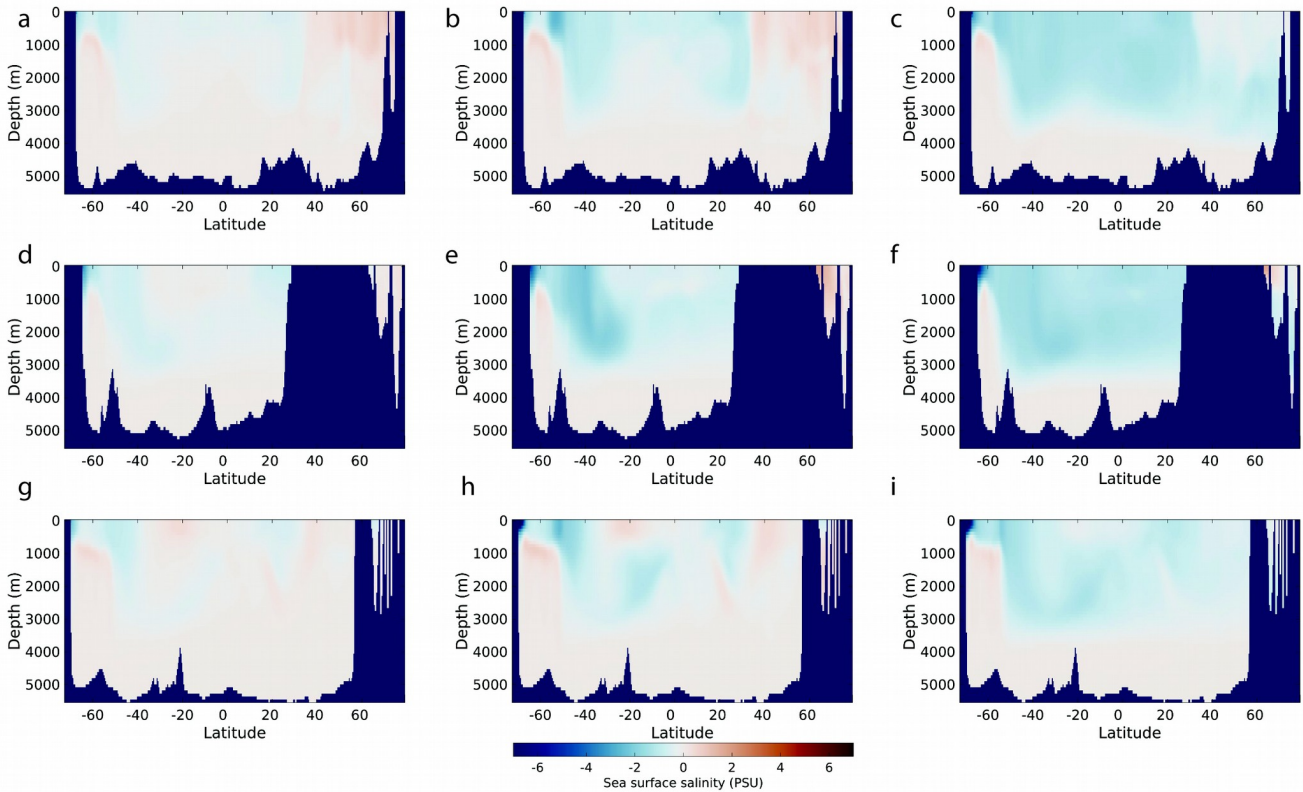

**Fig. S1. Salinity distribution at depth in RCP8.5FW.** (**A** to **C**) Salinity difference (RCP8.5FW minus RCP8.5CTRL) at depth, at longitude 342 in the Atlantic basin and decadally averaged for the time periods 2091-2100 (**A**), 2121-2130 (**B**), and 2191-2200 (**D**). (**D** to **F)**, The same but for the Indian Ocean at longitude 72. (**G** to **I**) The same for the Pacific Ocean at longitude 213.

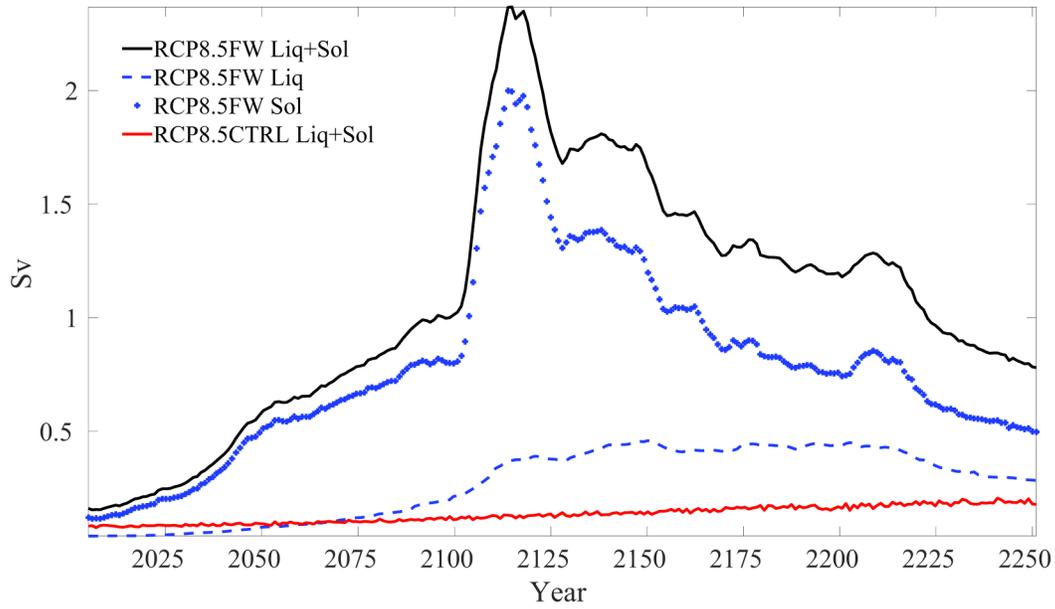

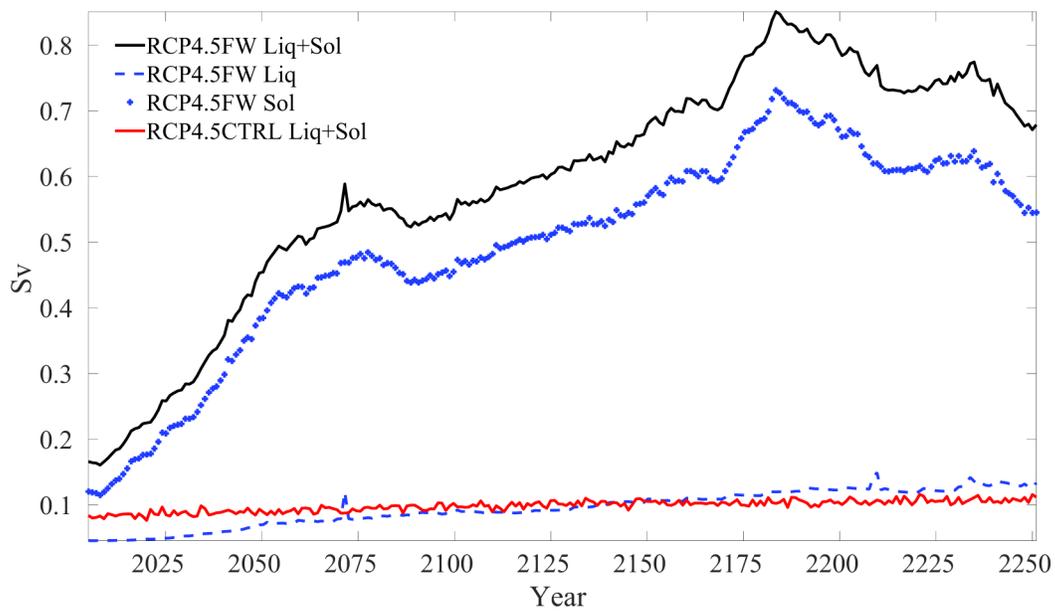

**Fig. S2. Freshwater forcing quantities.** (**A**) The freshwater forcing used in RCP8.5FW is shown with liquid and solid components separate, as well as combined, alongside the forcing computed by CESM in RCP8.5CTRL. (**B**) The same is (A), but for RCP4.5.

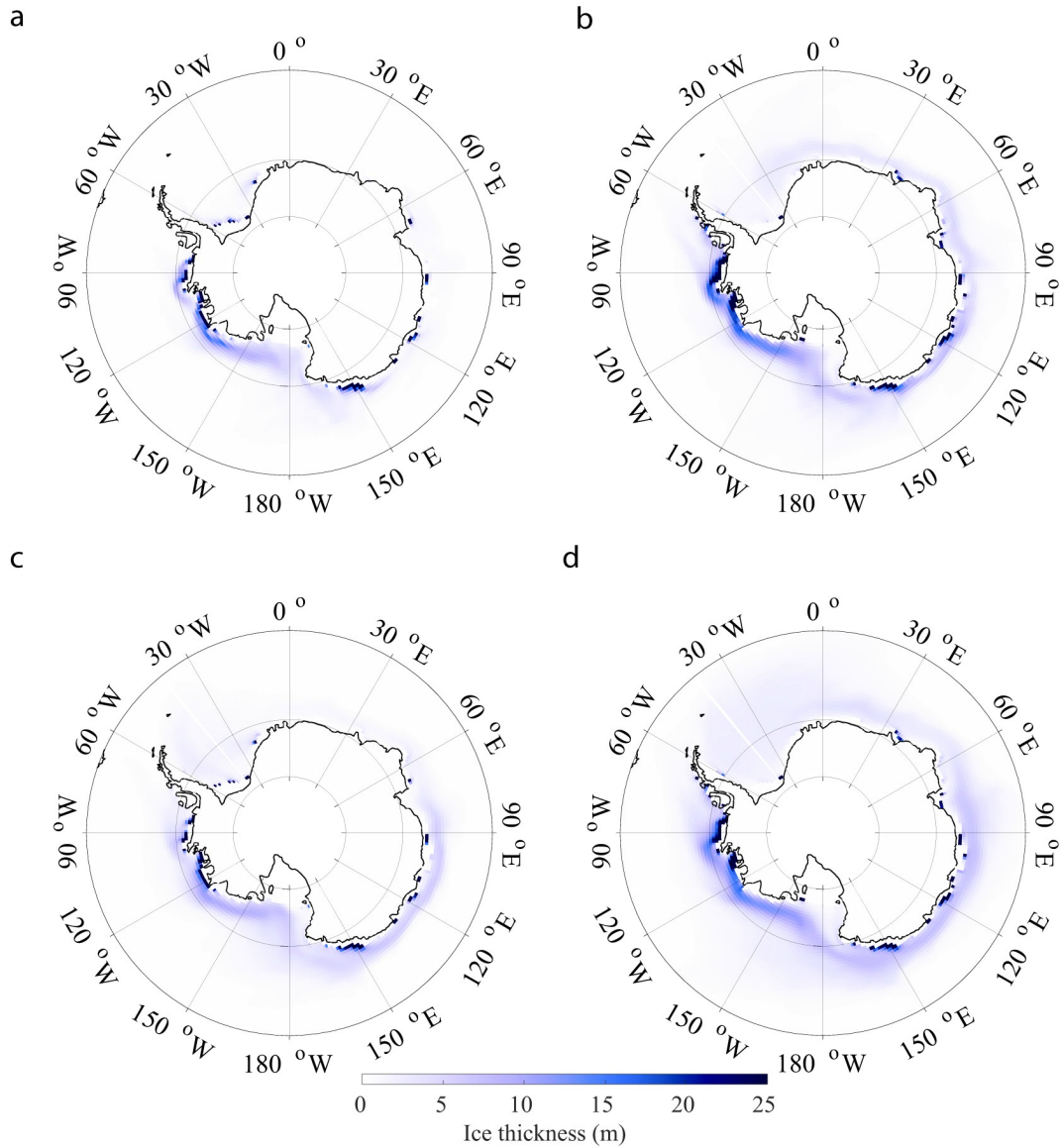

**Fig. S3. Southern Ocean sea ice in the 2190s.** (**A**) Southern Ocean sea ice in RCP8.5FW at the end of the 21st century decadally averaged from 2191-2200 for February. Grid cells where ice area is <10% and ice thickness is <0.005 m have been removed. (**B**) The same period is shown for RCP4.5FW. Note the more extensive sea-ice development for this time period compared to RCP8.5FW. (**C**) RCP8.5FW in September for the same time span as (A) and (B). (**D**) RCP4.5FW in September for the same time span as (A to C). RCP8.5CTRL RCP4.5CTRL are not included as there is virtually no ice in those runs for this time period (Fig. 2a).

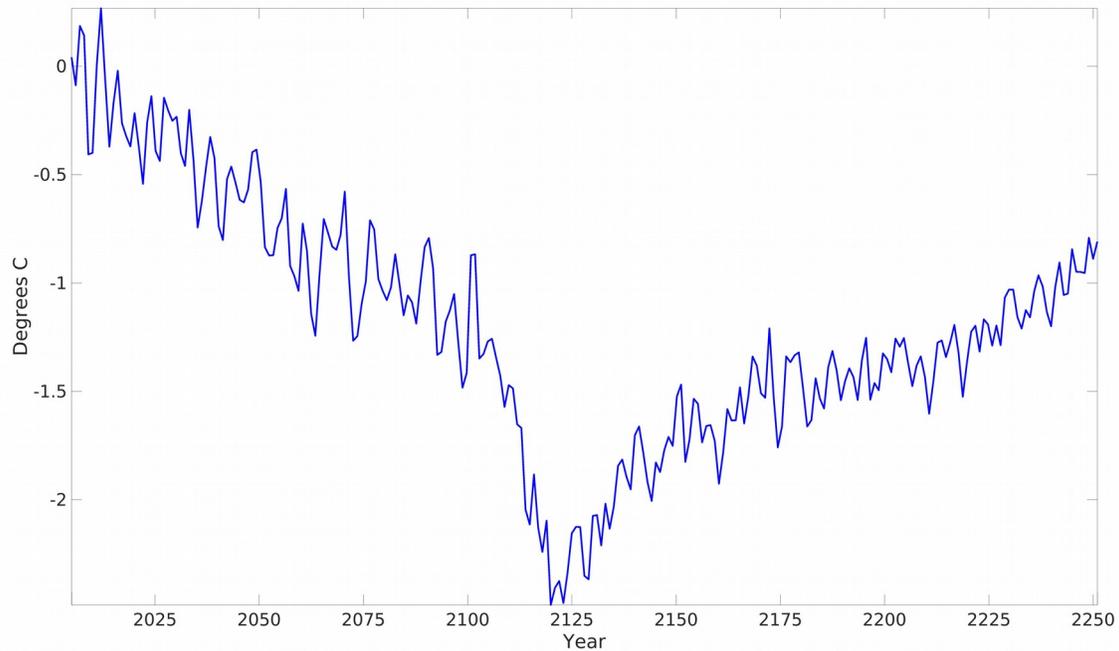

**Fig. S4. Globally averaged 2 meter air temperature anomaly.** The difference in 2 m air temperature between RCP8.5FW and RCP8.5CTRL is maximized during peak Antarctic ice loss, peaking at around 2.5°C between years 2120-2125. The freshwater perturbation delays the warming, but once the AIS is exhausted of ice the temperatures between the two runs begin to converge.

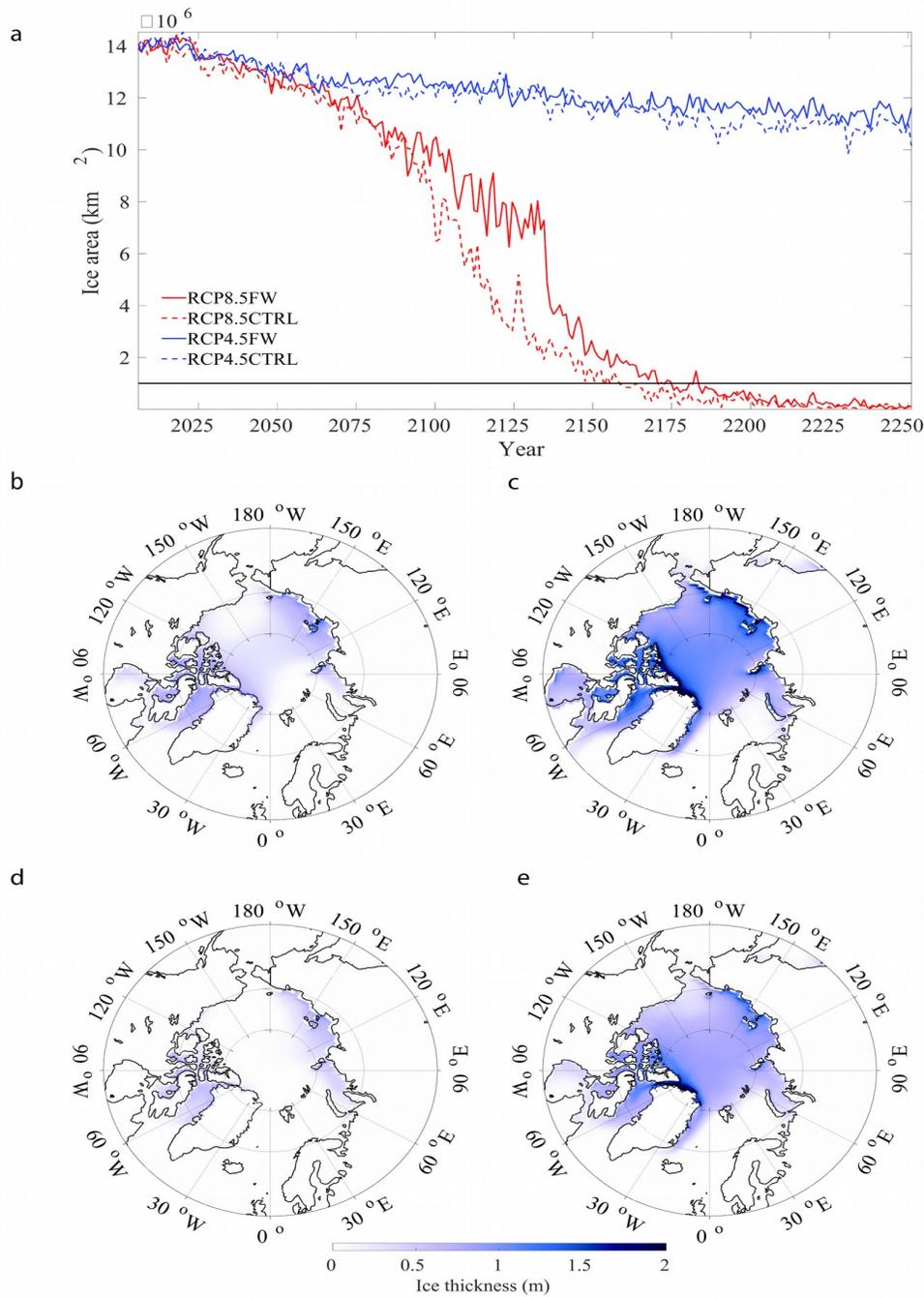

**Fig. S5. Winter Arctic sea ice.** (**A**) Arctic ice loss is delayed during the 21st century in RCP8.5FW due to delayed surface air temperature increases as a result of the Southern Ocean freshwater forcing. The black line represents ice free conditions defined as 1 million square kilometers. (**B**) RCP8.5FW, (**C**) RCP4.5FW, (**D**) RCP8.5CTRL, (**E**) RCP4.5CTRL show sea-ice thickness for February, decadally averaged from 2121-2130. Grid cells where ice area is less then 10% and ice thickness is less then 0.005 m have been removed.

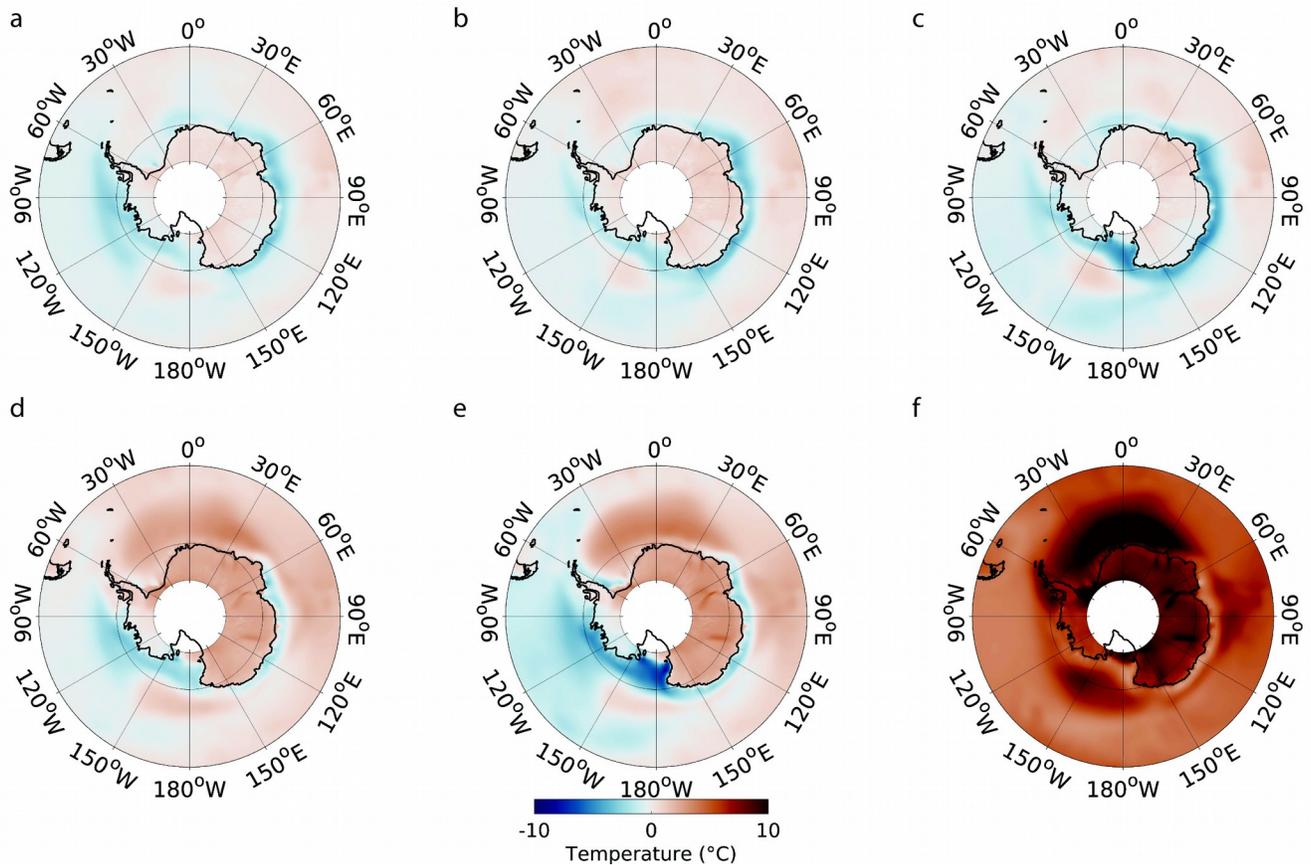

**Fig. S6. Southern Ocean 2m air temperature evolution.** (**A**) Surface air temperature for RCP4.5FW averaged from 2091-2100 minus the 2005-2014 average. The same is shown for (**B**) 2121-2130 and (**C**) 2191-2200. The expansion of sea ice where the freshwater perturbation was applied has lower SAT values than at the start of the run, due to the sustained freshwater forcing in this experiment. (**D** to **F**) The same time periods for RCP8.5FW shows that this effect is sustained only during freshwater discharge occurs; after that temperatures rise rapidly due to anthropogenic greenhouse gas forcing.

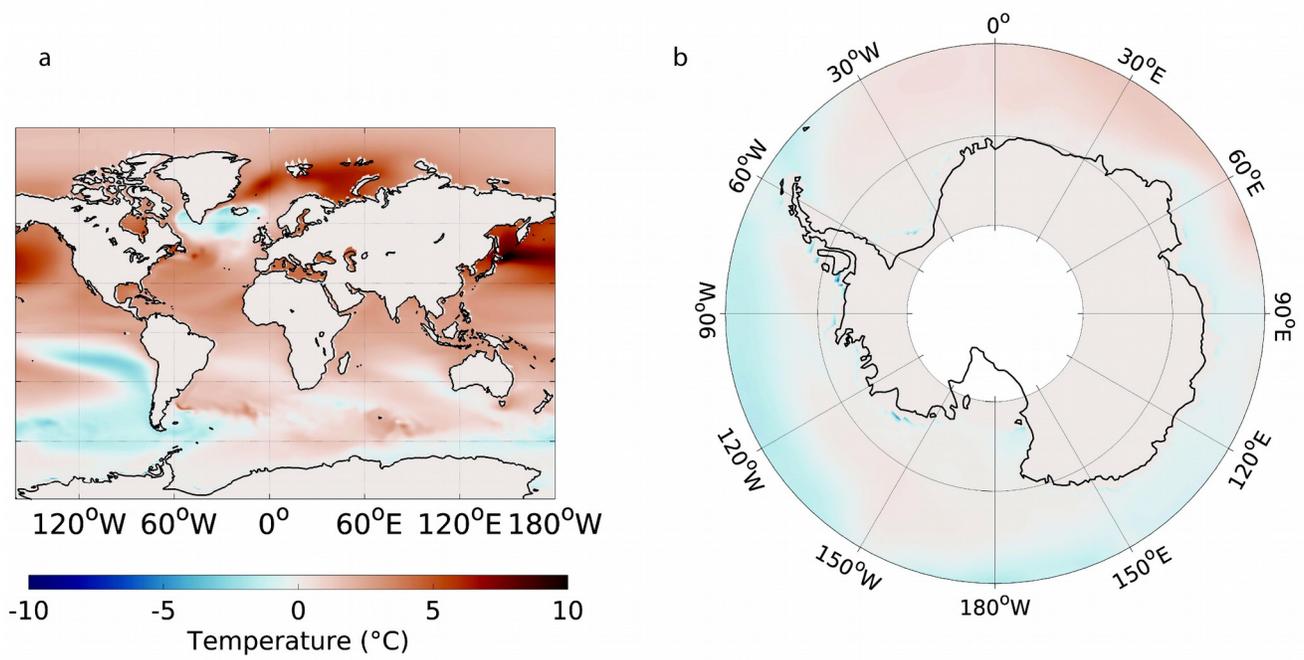

**Fig. S7. Sea surface temperature (SST) evolution.** (**A**) The SST values for RCP8.5FW decadally averaged from 2121-2130, compared to the decadal averages from 2005-2014 (first decade of the run) show that during peak freshwater forcing the SST values in the Southern Ocean are lower than at the start of the simulation. (**B**) The same data shown in a polar stereographic projection.

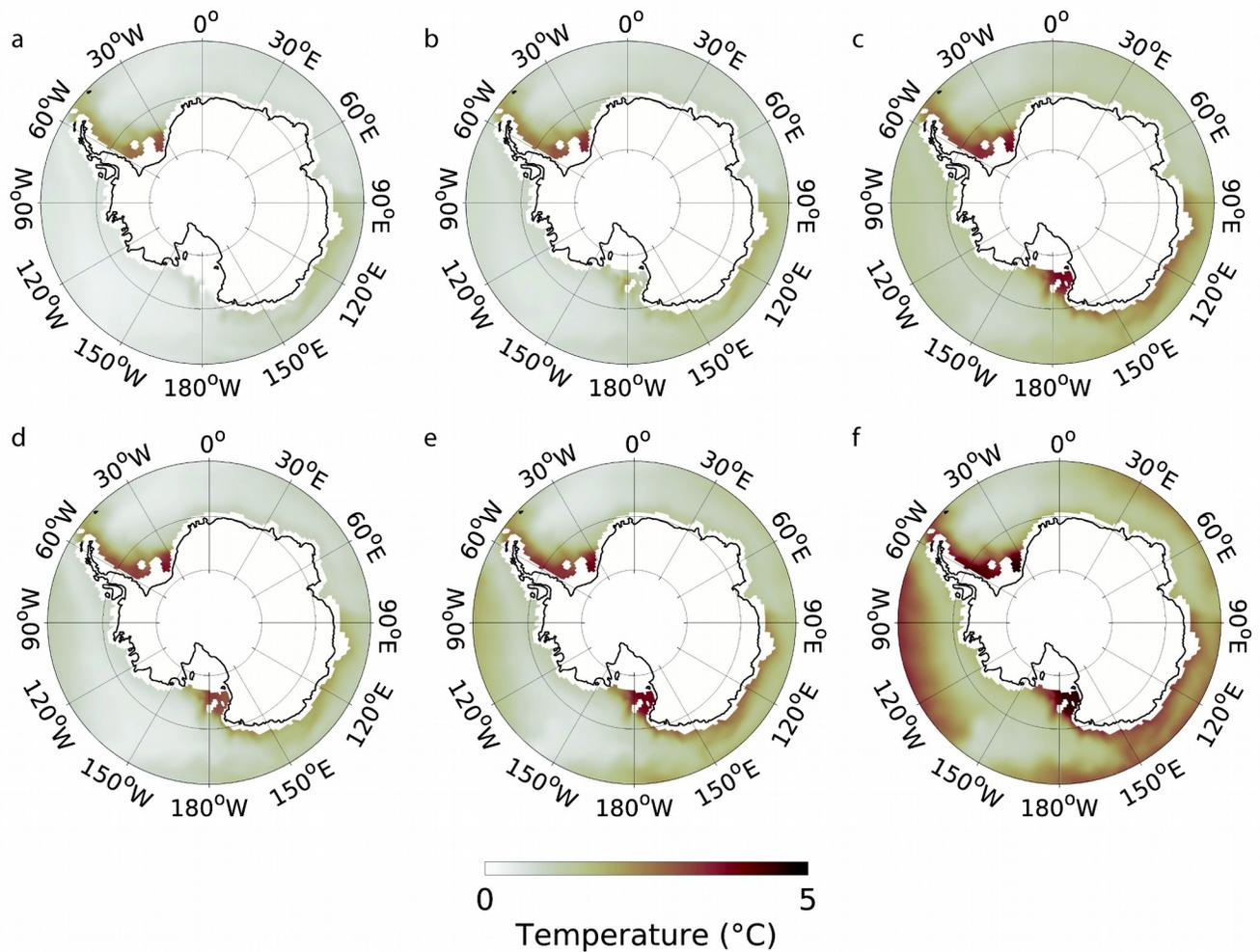

**Fig. S8. Ocean temperature evolution at 400 m in FW simulations.** (**A**) 400 m water temperature in RCP4.5FW with the 2005-2014 (first decade of the integration) average subtracted from the 2091-2100 average. (**B**) RCP4.5FW 2005-2014 average subtracted from the 2121-2130 average. (**C**) RCP4.5FW 2005-2014 average subtracted from the 2191-2200 average. (**D** to **F**) The same time periods as above but for RCP8.5FW.

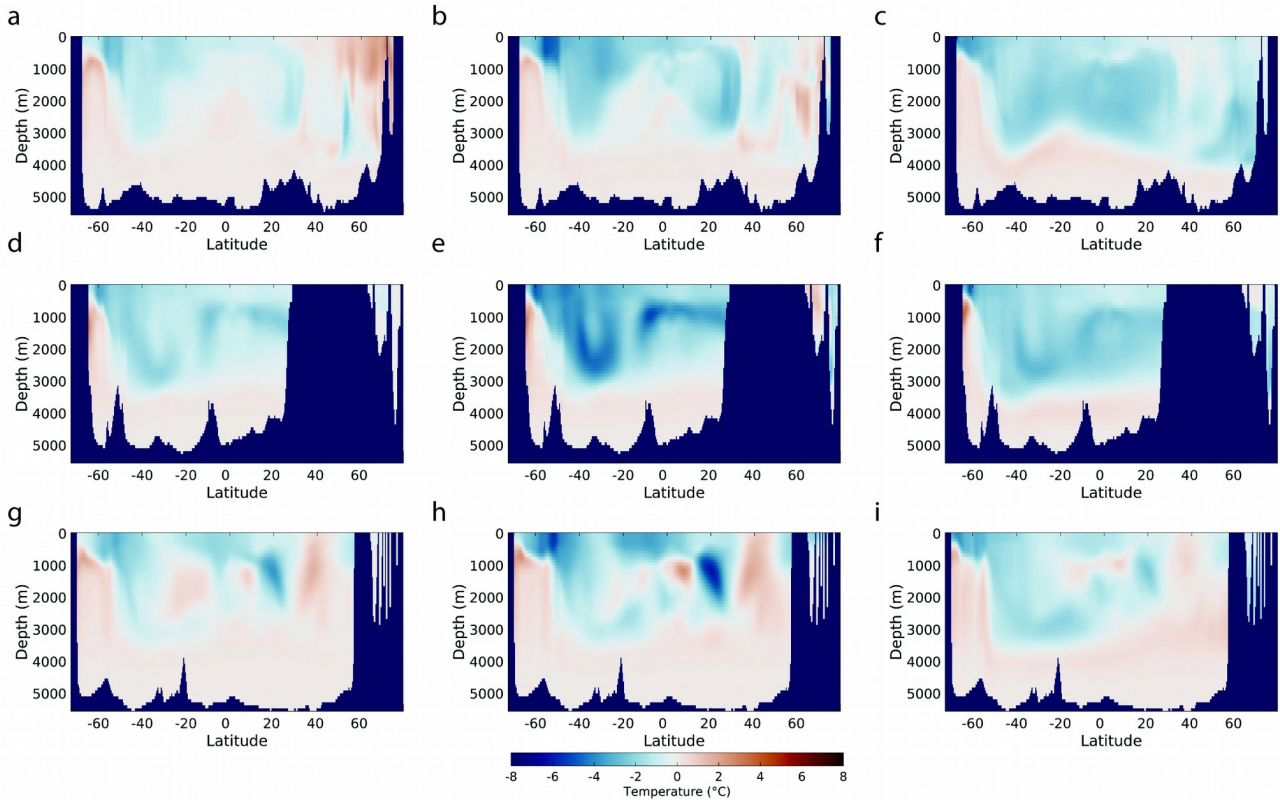

**Fig. S9. Temperature anomaly at depth.** (**A** to **C**) The temperature difference between RCP8.5FW and RCP8.5CTRL at depth for the Atlantic, decadally averaged for the time periods 2091-2100 (a), 2121-2130 (b), and 2191-2200 (c). (**D** to **F**) The same as (A to C) but for the Indian Ocean. (G to **I**) The same as (A to C) but for the Pacific Ocean. The cooler sub-surface ocean temperatures relative to the simulation without freshwater forcing are pervasive throughout much of the water column above 4000 m depth. Warmer relative temperatures in the perturbation run are evident at depths below 400 m in the Southern Ocean.

|                              | RCP8.5FW | RCP8.5CTRL | RCP4.5FW | RCP4.5CTRL |
|------------------------------|----------|------------|----------|------------|
| **2006**                     |          |            |          |            |
| GMST (C)                     | 14.0539  | 14.0120    | 14.1783  | 14.1226    |
| Increase over 1979-2000 (C)  | 0.3950   | 0.3531     | 0.5340   | 0.4637     |
| SO Temp                      | -19.7941 | -19.7299   | -19.7903 | -19.8114   |
| AMOC (Sv)                    | 24.9189  | 24.2989    | 24.3412  | 24.3237    |
| SO Ice Area ($10^6 km$)      | 3.1276e6 | 3.2739     | 2.8595   | 2.4876     |
| **2050**                     |          |            |          |            |
| GMST                         | 15.4120  | 15.7964    | 14.5183  | 15.3484    |
| Increase over 1979-2000      | 1.7531   | 2.1375     | 0.8253   | 1.6895     |
| SO Temp                      | -20.7809 | -18.0770   | -21.0767 | -18.3307   |
| AMOC                         | 19.4244  | 20.6395    | 21.6083  | 20.9936    |
| SO Ice Area                  | 5.4382   | 1.8630     | 5.4666   | 1.9562     |
| SO Ice Percent Change        | 73.8755  | -43.0960   | 91.1742  | -21.3617   |
| **2100**                     |          |            |          |            |
| GMST                         | 17.0289  | 18.4440    | 15.3697  | 16.1700    |
| Increase over 1979-2000      | 3.3700   | 4.7851     | 1.8161   | 2.5110     |
| SO Temp                      | -18.8967 | -14.1521   | -19.7687 | -17.5023   |
| AMOC                         | 15.2980  | 12.1330    | 19.9960  | 17.7662    |
| SO Ice Area                  | 4.2109   | 0.0837     | 5.0735   | 1.3677     |
| SO Ice Percent Change        | 34.6366  | -97.4443   | 77.4289  | -45.0207   |
| **2200**                     |          |            |          |            |
| GMST                         | 21.2402  | 22.5913    | 15.6360  | 16.9301    |
| Increase over 1979-2000      | 7.5813   | 8.9324     | 2.0693   | 3.2712     |
| SO Temp                      | -12.5720 | -7.9139    | -20.4821 | -16.4034   |
| AMOC                         | 8.3380   | 9.4018     | 19.6999  | 21.2097    |
| SO Ice Area                  | 1.3659   | 0.0000     | 5.7304   | 0.4388     |
| SO Ice Percent Change        | -56.3273 | -99.9993   | 100.4027 | -82.3600   |
| **2250**                     |          |            |          |            |
| GMST                         | 22.8242  | 23.6334    | 16.0015  | 17.3196    |
| Increase over 1979-2000      | 9.1653   | 9.9745     | 2.1852   | 3.6607     |
| SO Temp                      | -9.4375  | -6.3080    | -19.4973 | -15.6418   |
| AMOC                         | 7.7511   | 7.9471     | 19.2572  | 21.4680    |
| SO Ice Area                  | 0.4567   | 0.0000     | 4.2865   | 0.4293     |
| SO Ice Percent Change        | -85.3983 | -99.9999   | 49.9045  | -82.7427   |

**Table. S1. Select model values.** Tabulated model quantities include globally averaged 2 m surface air temperatures, 2 m surface air temperature rise averaged over 1979-2000, relative to (13.66°C) from the CESM pre-industrial simulation. 2 m air temperature averaged over the Southern Ocean, maximum AMOC strength in the North Atlantic, the area of the Southern Ocean covered by sea-ice, and the percent change in Southern Ocean area covered by sea-ice compared to the 2005-2014 average.